**Magnetotransport as diagnostic of spin reorientation: kagome ferromagnet as a case study**

Neeraj Kumar[1], Y. Soh[1],

1. Neutron and Mouns, Paul Scherrer Insitute, Villigen, 5232, Switzerland

Yihao Wang[2], Y. Xiong[2]

2. Anhui Province Key Laboratory of Condensed Matter Physics at Extreme Conditions, High Magnetic Field Laboratory of the Chinese Academy of Sciences, Hefei 230031, China

**Abstract:**

While in most ferro or antiferromagnetic materials there is a unique crystallographic direction, including crystallographically equivalent directions, in which the moments like to point due to spin-orbit coupling, in some, the direction of the spin reorients as a function of a certain physical parameter such as temperature, pressure etc. $Fe_3Sn_2$ is a kagome ferromagnet with an onset of ferromagnetism below 650 K, and undergoes a spin reorientation near 150 K. While it is known that the moments in $Fe_3Sn_2$ point perpendicular to the kagome plane at high temperatures and parallel to the kagome plane at low temperatures, how the distribution of the magnetic domains in the two different spin orientations evolve throughout the spin reorientation is not well known. Furthermore, while there have been various reports on the magnetotransport properties in the Hall configuration, the angular dependence of magnetoresistance has not been studied so far. In this paper, we have examined the spin reorientation by using anisotropic magnetoresistivity in detail, exploiting the dependence of the resistivity on the direction between magnetization and applied current. We are able to determine the distribution of the magnetic domains as a function of temperature between 360 K to 2 K and the reorientation transition to peak at 120 K. We discover that both out of plane and in plane phases coexist at temperatures around the spin reorientation, indicative of a first order transition. Although the volume of the magnetic domains in the different phases sharply changes at the spin reorientation transition, the electronic structure for a specific magnetization is not influenced by the spin reorientation. In contrast, we observe an electronic transition around 40 K, hitherto unreported, and reflected in both the zero-field resistivity and anisotropic resistivity.

**Introduction:**

Kagome $Fe_3Sn_2$ orders ferromagnetically below a Curie temperature of $T_C$ = 640 K based on SQUID magnetometry[1]. Previous studies using Mossbauer spectroscopy reported a Curie temperature of 612 K[2] and 657 K[3]. Below the ordering temperature, the easy axis of magnetization is parallel to the crystallographic c-axis. Initial study using Mossbauer spectroscopy noticed that a transition (SRT) occurs at 114 K[2]. Following studies using Mossbauer spectroscopy suggested that below 220 K, there are abrupt spin rotations occurring over a large temperature range 0-220 K with the spin direction close to the ab Kagome plane at low temperatures[3]. Further studies investigating the spin rotation using neutron diffraction combined with Mossbauer spectroscopy noted that the rotation is more complicated than a continuous rotation described by a unique angle or a simple abrupt rotation[4].

The spin reorientation transition (SRT) was recently revisited, using powder neutron diffraction, where the transition was suggested to occur over a large temperature range from 570 K to 75 K[1]. The order of the transition is not discussed in any of the previous reports and they report very broad transitions. Furthermore, previous reports measured powder sample, where the outcome could be representative of non-intrinsic features of the material. In the case of DyFe$_{11}$Ti, it has been very clearly shown that the sample quality can easily affect the sharpness of the SRT[5]. In this work, we investigate the magnetic behavior especially the SRT using high quality single crystals of Fe$_3$Sn$_2$. We use magneto-transport measurements as the main probe, and focus on the anisotropy in the resistivity due to the angle between magnetization and applied current. Anisotropic magnetoresistance (AMR) is a very useful tool to probe the bulk domain configuration, especially, if the magnetoresistance (MR) is not very high[6].

While there have been several reports on the large anomalous Hall effect in Fe$_3$Sn$_2$,[7, 8], MR in Fe$_3$Sn$_2$ has not been investigated in detail so far, except for MR for H//c [8, 9]. We investigate in detail the temperature dependence of MR for fields along different directions and discuss the various contributions. Using AMR, we examine the SRT in detail and unambiguously demonstrate the first order nature of the transition. We find the transition range to be much narrower 90 K-140 K than previous reports, with the transition peaked at 120 K. We speculate the reason for the broad transition observed in previous studies to be due to the samples being in powder form. Our magneto resistivity data is compared against bulk magnetization measurements to understand the influence of the magnetization state on the resistance.

In addition to the SRT, we discovered that there is an electronic transition at T = 40 K based on AMR and zero field resistivity data. The origin of this transition is not clear to us at present.

**Experimental details:**

Crystals were grown using vapor transport. As-grown crystals are in the form of thin platelets and carry the hexagonal crystal structure morphology. Structure of the crystal is confirmed using single crystal XRD and Laue diffraction. A crystal of lateral dimensions 1 mm X 0.5 mm and thickness of 35 μm was selected for the measurements. Magnetization measurements were carried out using a Quantum Design SQUID VSM. Magneto-transport measurements were conducted using a Quantum Design PPMS with the option to rotate the sample w.r.t. the magnetic field in-situ. The long side of the crystal was found to be parallel to crystallographic a axis and this direction was selected for applying current. Magneto-transport measurement was performed in three configurations: (i) parallel configuration, i.e., magnetic field was applied parallel to the current along a (ii) transverse configuration, with magnetic field in the kagome plane but perpendicular to the current, we denote this direction as $a_\perp$, and (iii) out of plane c configuration. Magnetization was also measured for these directions. In-plane directions are shown in Fig. 1(c) inset.

**Results and discussion:**

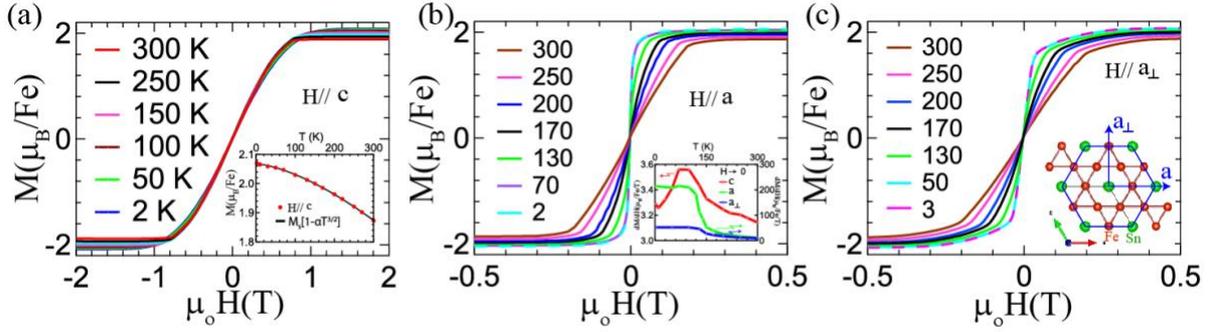

Figure 1: Magnetization (M) for Field (H) parallel to (a) c axis, (b) a axis, (c) $a_\perp$ axis. Inset (a) the saturation M as a function of temperature. Inset (b) Susceptibility for field along each of the 3 directions in the limit of zero field. Inset (c) Kagome plane together with the directions along which field has been applied in the plane.

The bottom inset in Fig.2 (a) shows the crystal structure of $Fe_3Sn_2$ with kagome plane shown in Fig. 1(c) inset. Hexagonal bilayers of $Fe_3Sn$ in the ab-plane are stacked along the c-axis, separated by Sn hexagonal layers[10]. We examine the bulk magnetometry as shown in Fig. 1. As $Fe_3Sn_2$ is a soft ferromagnet, no hysteresis is observed in the magnetization. For H//a, a clear change in saturation field and slope of M vs H curve is seen upon lowering the temperature, confirming the SRT. Similar behavior for H//$a_\perp$ is also seen which shows an additional turn in M vs H curve and higher saturation field. A higher saturation field could be either due to anisotropy energy or due to the demagnetization factor because the sample geometry is elongated along a, or both. As seen in Fig. 1(a), for H//c, change in the MH curves is very small, which is due to the masking of changes in saturation field by the much larger demagnetization field. The inset in Fig. 1(c) shows the susceptibility dM/dH at zero magnetic field as a function of temperature for magnetic field applied along each of the three directions as derived from the M vs H data in Fig. 5. A clear change in the susceptibility along a and $a_\perp$ beginning at 150 K, support the SRT. Based on the susceptibility data, we estimate that the SRT is completed at a temperature above 90 K. For H//c, change in the susceptibility is very small, primarily again due to demagnetization field. Fig. 1(a) inset shows the saturation magnetization as a function of magnetic field for H//c. The magnetization follows Bloch's law $M = M_S(1 - \alpha T^{\frac{3}{2}})$.

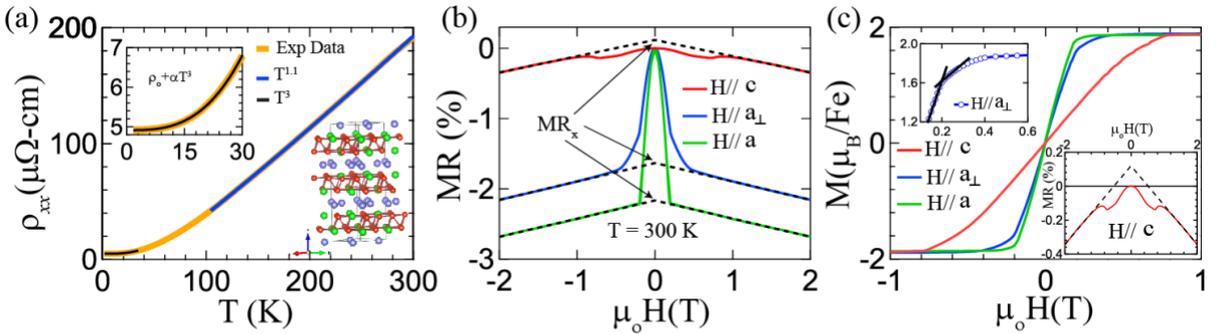

Figure 2: (a) Resistivity vs temperature shows typically metalic behavior with different fitting. Inset shows the crystal structure of $Fe_3Sn_2$ showing stacking of kagome bilayers along c-axis. Upper inset highlights the fitting below 30 K. (b) Magnetoresistance for field along a, $a_\perp$ and c directions. (c) Magnetization for these directions. Top inset shows the turn in the M for H//$a_\perp$. Bottom inset highlights the magnetoresistance for H//c.

The resistivity of $Fe_3Sn_2$ shows a typical metallic behavior with a very high residual resistivity ratio (defined as $\frac{\rho(300K)}{\rho(2K)}$) of 40 (Fig. 2(a)). The resistivity value at room temperature is around 200 μΩ-cm and agrees well with other recent reports [8, 9]. The resistivity is almost linear above 100 K, but can be more appropriately described by a $T^{1.2}$ dependence. At low temperature below 30 K, the resistivity varies as $T^3$. A $T^3$ dependence could indicate s-d electron scattering as the possible predominant mechanism at low temperature [11]. As we will describe later, it is possible that a phase transition occurs around 30 K.

Fig. 2(b) shows the MR (defined as $\frac{\rho(H)-\rho(0)}{\rho(0)}$) of Fe$_3$Sn$_2$ at 300 K for a magnetic field applied in the ab-plane along longitudinal, transverse directions, and along out of plane direction (c-axis). For each configuration, a negative MR is observed, which is linear above a certain saturation field. The saturation field for each direction reflected in the MR is in accordance with the corresponding saturation field in the magnetization as shown in Fig. 1(c). Correspondingly, this means that the MR features below the saturation field reflect changes in the magnetic domain configuration on application of magnetic field. It is to be noted that at low temperature, the MR becomes ultimately positive at high fields after the low field negative MR, in contrast to the behavior at 300 K. That positive MR is not associated with the magnetization and magnetic domains and is shown in Fig. 3. (see Fig. S1 in the supplementary section).

The isotropic magnon scattering contribution persists down to nearly 150 K as shown in Fig. 3(a) as a negative slope of the MR above the saturation field. This contribution starts to decrease below 150 K as magnons freeze. The Lorentz MR is normally negligible at high temperature and starts to become significant only at low temperatures around 100 K. The MR behavior at 100 K (Fig. 3(a) inset) shows a clear competition between the two effects and can be easily fitted to polynomial $\rho(H) = \alpha H + \beta H^2$. A similar behavior has been seen in other ferromagnetic materials as well [12]. As the temperature decreases, the linear negative MR is no longer significant. Below 60 K, the MR is completely positive and better described by a power law $\rho(H) = \rho_o(1 + \alpha H^p)$ (Fig. 3(b)). The exponent of the field is close to 1.8 at 60 K and to 1.3 at 2 K. As noted in Ref [13], various semi metallic materials are seen to exhibit positive MR with varying exponents. The polynomial $\rho(H) = \alpha H + \beta H^2$ also can be used to fit the MR, but the power law fitting is relatively better (see supplementary section). At 2 K, positive MR is seen for field along the three directions (Fig. 3(c)) studied in this paper, however with different exponent. For current and field parallel to each other, the MR at 2K is almost linear. Causes for the sub-quadratic MR has been ascribed to either the dependence of the mobility on the magnetic field[14] or due to the itinerant carriers encountering sharp cornered surfaces[15].

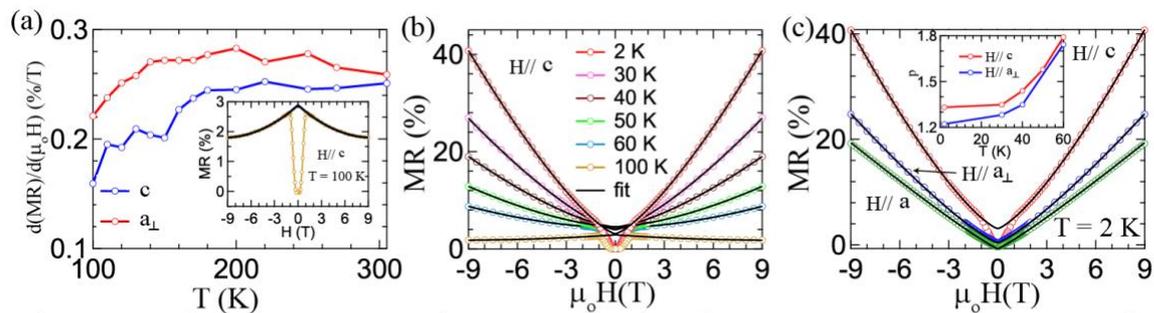

Figure 3: (a) Electron magnon scattering term as a function of temperature. Inset: Magnetoresistance for H//c at 100 K. (b) Magnetoresistance at various temperatures below 100 K for H//c axis showing excellent fitting with a power law $\rho=\rho_o(1+\alpha H^p)$ (except at 100 K). (c) Magnetoresistance at 2 K for field along each of the three directions. Inset shows the change in exponent p of the power law fitting.

We now turn our attention to the low field MR. The resistivity of a fully magnetized system in the limit of zero external field $\rho_x$, where x refers to the direction of the magnetization, or the corresponding magnetoresistance $MR_x = (\rho_x - \rho(0))/\rho(0)$ is obtained by extrapolating the data at high field to zero field as shown in Fig. 1(b). $MR_x$ reflects the anisotropic resistivity, which depends on the direction of the magnetization and arises from spin-orbit coupling. A positive value for $MR_c$ and a negative one for $MR_a$ and $MR_{a_\perp}$ is obtained. As the current is applied along [a], we define the anisotropic magnetoresistance (AMR) ratio as $MR_a - MR_c$. An AMR ratio of -2.2% is observed. A negative AMR is opposite to the conventional behavior of ferromagnets. However, it is seen in systems such as $Fe_3O_4$ [16] and 2D ferromagnets such as $Cr_2Ge_2Te_6$[17] and $Fe_3GeTe_2$[18]. Conduction by the minority spin carriers has been suggested as the reason behind a negative AMR[19].

$MR_c$ at 300 K is much smaller than the corresponding values for the other magnetization directions, which is in line with c being the easy axis at room temperature. However, for all the magnetic domains to be along c, $MR_c$ should have been zero, as both parallel and antiparallel domains contribute equally to the resistivity. A non-zero $MR_c$ means that, (i) either the magnetization axis is oriented close to but not exactly parallel to the c-axis or (ii), the sample consist of domains with multiple orientations with domains along c being the significantly larger fraction. As we will show later, the latter is more likely to be the case. We estimate the out of plane domain fraction using the zero field anisotropic resistivity values $\rho_x$. Assuming an effective medium model

$$(1-x)\rho_c + \frac{x(\rho_a + \rho_{a_\perp})}{2} = \rho(0),$$

where $x$ is the volume fraction of in-plane magnetic domains and the two in-plane directions are averaged to represent the in-plane resistivity. At 300 K, $x$ is found to be 0.08. We apply this concept later to extract the temperature dependence of the distribution of magnetic domains and show that the assumption of having a volume distribution of magnetization with different magnetic anisotropy is justified based on our MR data.

For H//a, and H//$a_\perp$, the large negative MR at 300 K below the saturation field is due to the rotation of the magnetization from the high resistivity M//c state to the low resistivity M//ab plane configuration. For H//$a_\perp$, however, there is an extra change in slope before saturation and an overall higher saturation field. This is because $a_\perp$ is not a high symmetry axis. When the field is applied along $a_\perp$, the magnetic domains first align toward a high symmetry axis, and later eventually toward $a_\perp$ as the field becomes stronger. This behavior near saturation is also seen in the magnetization data as shown in Fig.1(c).

As shown in the bottom inset of Fig. 2(c), for H//c, the MR is non-monotonic below the saturation field. Up to 0.7 T, the MR decreases with the increasing field and then after that increases rapidly up to 1 T. Such behavior is can be explained by the rotation of in-plane domains on application of field in the presence of a tilted secondary easy axis at a small angle with respect to the ab-plane exits. The in-plane fraction of domains will rotate toward this easy axis at low field if the anisotropy energy is lower as compared to anisotropy energy for out of plane rotation. At higher field, domains will eventually rotate toward c axis causing increase in the MR[20]. For comparatively lower resistivity for the secondary axis direction, which is possible due to in-plane anisotropy as seen in Fig.2 (b), a minima in resistivity for H//c will occur.

Similar MR behavior for $Fe_3Sn_2$ was seen previously [21], where the field values at the lowest MR and the highest MR were explained as the transition points from bubble to skyrmionic bubble and further to ferromagnetism phase, respectively. While it could be possible that there is an accompanied change of magnetic phase, we believe that the main drive for the trend in the MR is the magnetic domain behavior as explained above. Further, we believe that this small volume of in-plane domains arises from a competition between in plane and out of plane magnetic anisotropy, which facilitates the formation of skyrmionic bubbles. Above the saturation field, the MR is linear in all the cases and is caused by the suppression of electron-magnon scattering on application of a magnetic field as discussed previously.[12]

To further understand the magnetic domain composition, we explore the temperature dependence of the low field MR. Fig. 4(a) and (b) show the MR for H// a at various temperatures between 360 K and 2 K. As seen in Fig. 3(a), as the temperature decreases from 360 K to 140 K, the negative $MR_x$ increases in magnitude from 2% at 360 K up to 4% at 140 K. Below 140 K, there is a decrease in the magnitude of the $MR_x$ with decreasing temperature. $MR_x$ decreases to 0.3% at 2K. The saturation field monotonically decreases with the decreasing temperature in accordance to the magnetization data.

Fig. 4(c,d) shows the corresponding data for $H//a_\perp$. Similar to the above, the magnitude of the negative MR increases with decreasing temperature up to 150 K, and decreases after that. However, in this case, the AMR becomes positive below 80 K. For H//c, where the AMR is positive at 300 K, the positive AMR becomes more positive with decreasing temperature. These observations are compiled in Fig. 5(a), which shows the temperature dependence of the AMR between 2 and 300 K. While in most of the cases, the AMR is estimated by a linear extrapolation of the data above saturation as shown in Fig. 2(b), at low temperatures a power law is used to extrapolate the positive MR to zero field.

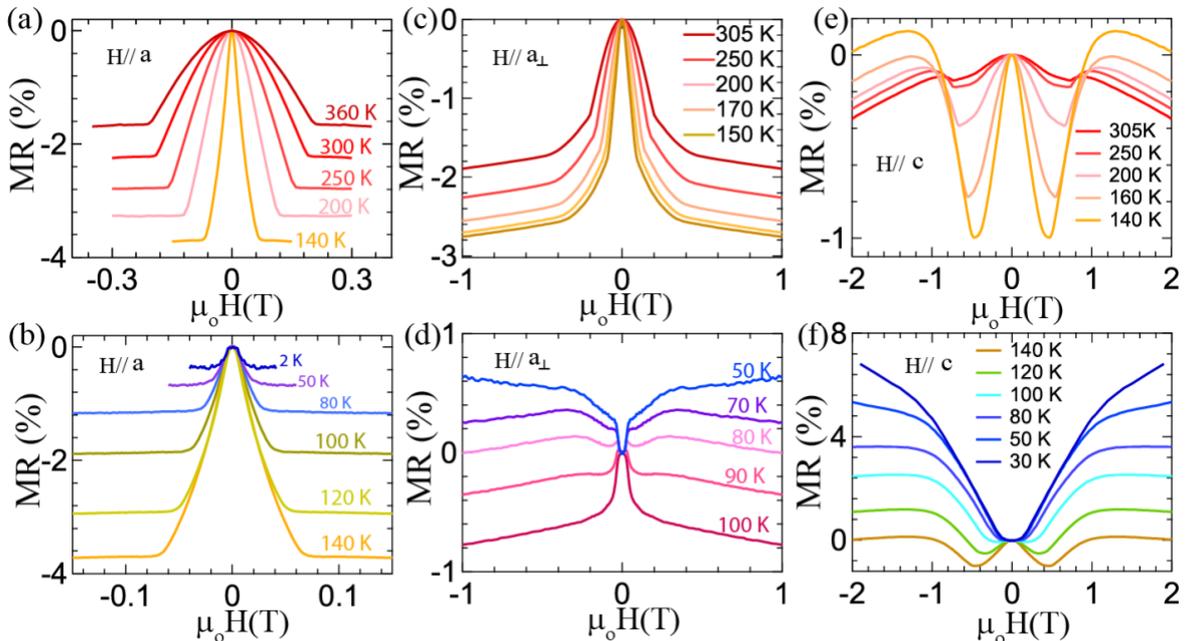

Figure 4: Magnetoresistance at several temperatures for magnetic field along (a,b) a, (c,d) $a_\perp$, (e,f) c.

As shown in Fig. 5(a), at 150 K, the trend of the MR changes for all three directions, suggesting the spin reorientation transition as the common origin, where the majority of the domains reorient from the c axis into the ab-plane. Using the same effective medium model, as shown in Eq. 1, the volume fraction of the in-plane magnetic domains is calculated and shown in Fig. 5(b) at temperatures ranging from 305 K down to 2 K. The volume fraction curve clearly reveals the SRT. From 300 K toward 150 K, there is a slow increase in in-plane volume fraction, followed by a rapid increase below 150 K. At 80

K, 90 % volume fraction is magnetized in the ab-plane. At 70 K, roughly the entire sample is magnetized in the ab-plane. By taking a derivate of the curve, the SRT temperature is found to be T=120 K. It is to be noted that an assumption regarding the equal population of domains in a and $a_\perp$ direction is made, which likely is not strictly true since a is a principal axis but $a_\perp$ is not. Nevertheless, it is sufficient to provide a good estimate of the transition behavior.

We can qualitatively explain the temperature dependence of the low field MR and AMR based on the SRT and volume fraction of magnetic domains. From 300 K down to 150 K, the majority of the domains are magnetized parallel to the c-axis. Therefore, in the parallel and transverse configuration, the application of a magnetic field will force these domains to rotate its magnetization to the ab plane leading to a lower resistivity and correspondingly to a negative AMR, whereas there would be little AMR for the out-of plane configuration. On the other hand, as the volume fraction of in-plane domains start increasing below 150 K, the application of a magnetic field in the parallel and transverse configuration would have a smaller effect than before, resulting in the decrease of the magnitude of the negative AMR causing the reversal of the trend at 150 K. For the out-of-plane configuration, the application of a magnetic field when the volume fraction of in-plane domain is significant will increase the positive AMR as the in-plane domain volume increases. The sign change from negative to positive AMR around 80 K in the transverse configuration can be understood based on the completion of the SRT with the magnetic easy axis being along directions equivalent to a. Above 80 K, there is still some volume fraction with magnetic domains along [c]. Therefore, the MR and AMR is negative. However, once the SRT is complete with the magnetic domains being magnetized along directions equivalent to a, the zero-field resistivity is lower than the anisotropic resistivity in the transverse configuration, giving rise to a positive MR and AMR.

Coming back to the butterfly MR for H//c, the magnitude of the negative MR reflected in the minima increases with decreasing temperature up to 150 K, which is consistent with slowly increasing ab-plane domain volume. The decrease in magnitude of this MR minimum below 150 K is again consistent with the SRT, where, although the ab-plane domain volume increases, the MR associated with ab domains decreases rapidly.

Comparing the data in Fig. 4(a) and (b), we observe a qualitative change in the shape of the MR vs H curve from a dome to a tower shape at low temperature. We argue that this change in shape is associated with the different magnetic domain configuration. The derivative of the MR, as shown in Fig. 5(c) further emphasizes this point, where at high temperatures the derivative peaks at a higher field and at low temperatures at lower fields. This difference is due to the difference in the anisotropy energy of c domains, which are majority domains at high temperature and ab-plane domains not parallel to a, which are the cause of MR and AMR at low temperature. The derivative curves for the intermediate temperature range such as at 150 K and 200 K, reflect a superposition of two peak shapes described above. This shows that the SRT does not occur by a continuous gradual rotation of M in the whole sample as in a second order phase transition but rather through a co-existence of both out of plane and in-plane domains whose volume fraction changes as the SRT occurs. The in-plane domain volume increases as the out-of-plane domain volume decreases as the system is cooled. This coexistence is the hallmark of a first order phase transition and validates our assumption of associating a volume fraction and AMR for each magnetization direction to describe the measured resistivity at a given temperature.

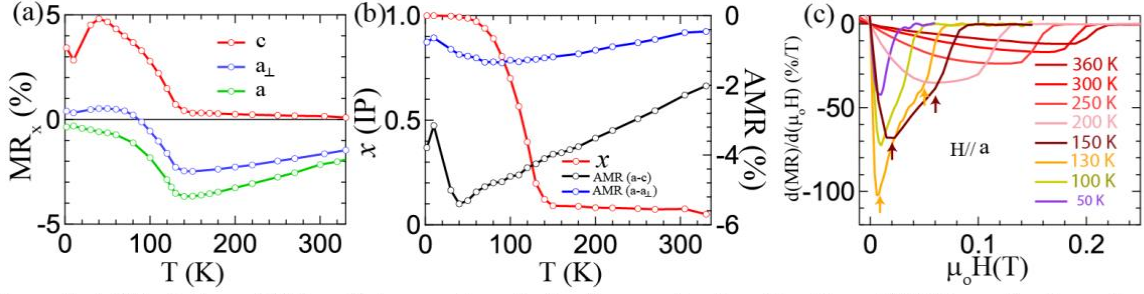

Figure 5: (a) Evolution of $MR_x$ with temperature for the 3 magnetization directions. (b) Volume fraction of in-plane domains $x$ is shown on left axis. Right axis shows the net AMR for two different configurations. (c) First derivate of Magnetoresistance for H//a clearly highlight the change in the easy axis and coexistence of two phases at some of thetemperatures.

We now turn our attention to the magnitude of the AMR ratio. As seen in Fig. 3(b), the AMR does not show any visible change at the SRT. Rather, at low temperature around 40 K, we observe a decrease in the magnitude of the AMR between $\rho_c$ and $\rho_a$ which seems to be mainly driven by the decrease in $\rho_c$. The reason behind this decrease is not clear. The zero-field resistivity also displayed a unique temperature dependence below 40 K. We speculate that both transitions have the same origin. For the AMR ratio between a and $a_\perp$, a similar trend is seen, however their extremum is seen around 80 K. Further experiments will be needed to discern the source of these anomalies in the data.

**Conclusion:**

We have been able to gain insight about the spin reorientation in $Fe_3Sn_2$ such as the distribution of the magnetic domains in the high vs low temperature phase, the transition temperature as the system undergoes a spin reorientation, and evidence for phase coexistence by probing the magnetoresistance. Our results rule out the scenario of a continuous rotation of the magnetic easy axis and are consistent with previous report describing a first order phase transition by SQUID magnetometry and MFM[22]. A resistivity probe, which is not a direct probe of magnetization, turns out to be complimentary to the magnetometry probe and reveals information about the magnetic property of the system not readily available through magnetometry measurements. In the case of bulk magnetometry of a soft magnet, there is essentially no remnant magnetization at zero external field since magnetic domains of opposite magnetization cancel each other. On the other hand, when probing resistivity, domains with opposite magnetization do not cancel each other, and therefore the zero-field resistivity contains information about the different magnetization directions such as M//c vs M//a. In order to probe the magnetization using magnetometry an imbalance of up and down magnetic domains needs to be created resulting in measurements away from zero field and smearing the phase transitions. Therefore, a careful application of magnetoresistance can be a powerful tool for probing the magnetization as we have shown in this paper.

**Supplementary section:**

Section 1:

The MR for H//[a] at 2 K and field upto 9 T is shown. The MR is negative at very small field ~0.05T as discussed in the main text. Above this, the MR is positive and increases almost linearly.

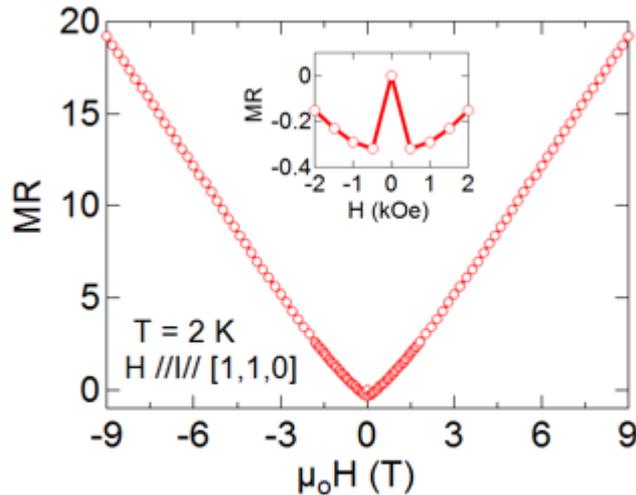

Figure S1: MR for H//a at 2 K show positive resistivity. Inset reveals a small negative MR at very low field.

Section 2:

Details on estimation of resistivity at zero field:
1. Above 100 K, the MR is overall negative and MR vs H curve is linear above the saturation field. This linear behavior is due to the suppression of magnon scattering. Thus, a linear fit is enough to estimate the resistivity at zero field.

2. Below 100 K, due to the semimetalic behavior, a positive MR contribution is seen, which is often non-linear. A power law $\rho(H) = \rho_o(1 + \alpha H^p)$ gives a reasonable fit. Other fitting functions such as polynomial $\rho(H) = \alpha H + \beta H^2$ also give good quality fits. The extrapolated values are not affected by the fitting method used. The figure below shows the residue in MR of the two fits at 2 K for H//c. The coefficients of a polynomial fitting are shown in the next figure.

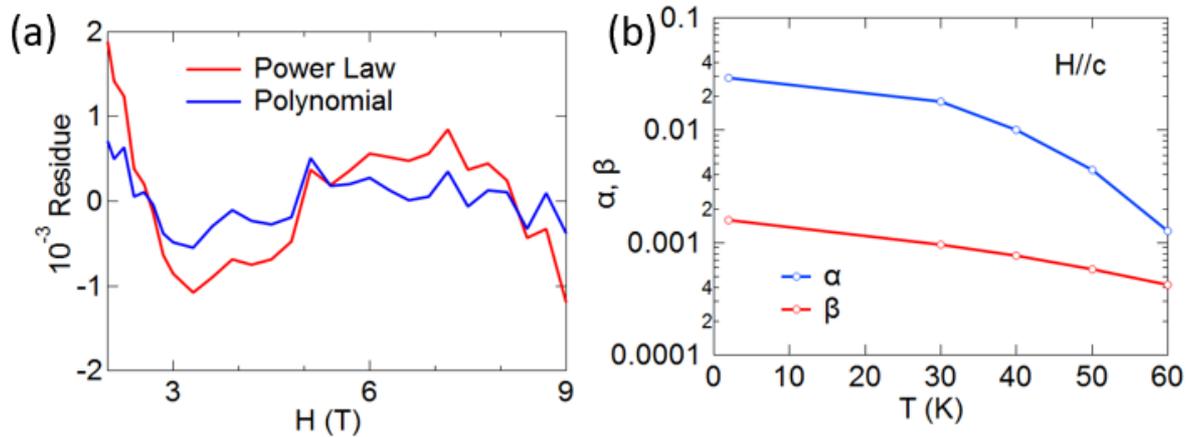

Figure S2: (a) Residue of power law and polynomial fitting for H//c at 2 K. (b) For H//c, the variation of coefficients in the polynomial fitting.

Section 3:

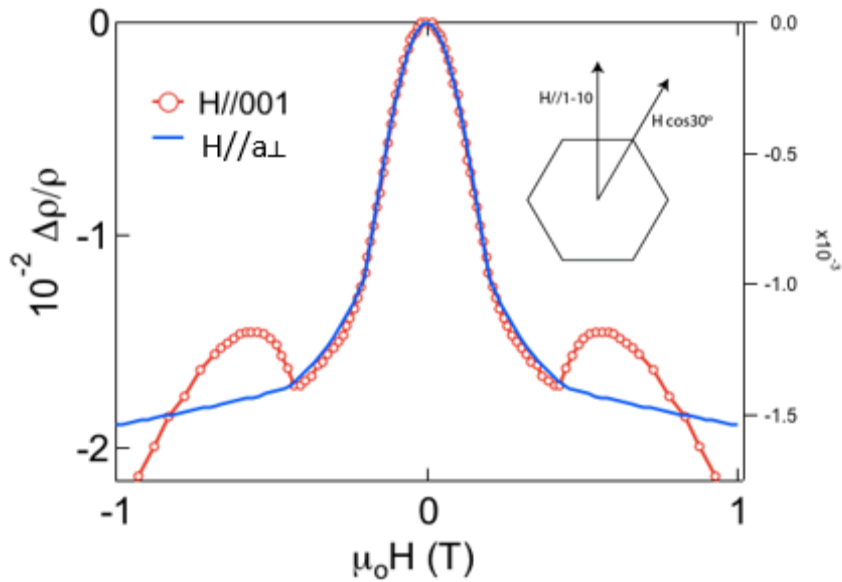

Figure S3: Comparison of MR for H//c and H//a⊥.

The MR along c and a are compared to show the similarity in their behavior. The magnetic field and net MR scale for H//c are adjusted to match the two curves. The change in field scale is necessitated due to the different demagnetization factor depending on the measurement configuration. In both the cases, MR first decreases at a rapid pace, and subsequently at a slower pace. This mechanism could be due to a secondary easy axis as shown in the figure. For the H//a case, when magnetic field is applied, an effective magnetic field Hcos30, i.e. 0.9H is applied along a principle axis.